\documentstyle[12pt]{article}
\begin{document}
\begin{center}
Chaos control in traffic flow models\\
       Elman Mohammed Shahverdiev,\footnote{e-mail:shahverdiev@lan.ab.az\\
On leave from Institute of Physics, 33, H.Javid avenue, Baku 370143, 
Azerbaijan}\\
Department of Information Science, Saga University, Saga 840, Japan\\
e-mail:elman@ai.is.saga-u.ac.jp\\
           Shin-ichi Tadaki\\
Department of Information Science, Saga University, Saga 840, Japan\\
e-mail:tadaki@ai.is.saga-u.ac.jp\\
\end{center}
Nowadays traffic flow problems has acquired interdisciplinary 
status due to the following reasons:first of all to investigate 
traffic flow models methods from different branches of science such as 
hydrodynamics, theory of magnetizm, cybernetics, etc. are applied;
secondly results of investigations of traffic problems with different
approaches could be adequate in various scientific directions, see, 
e.g. refs.1-9.\\
Mean field approach is one of widely used approaches in traffic 
problems.(for some publications, see, ref.8-10 and references therein).
It is well-known that nowadays cellular automaton (CA) models has 
extensive applications to the traffic flow models.In this paper we dwell
on just two models from the mean field theory:one- and two-dimensional
systems.First we will consider one dimensional model.\\
Recently, the authors of ref.8 have presented microscopic derivations of 
mean field theories for CA models of traffic flow in one dimension.They 
established the following mapping between the average velocities of cars $v$ 
at times t+1 and t:
$$v(t+1)=(2-f)v(t)-(1-p)^{-1}v^{2}(t)+p(1-p)^{-1}v^{3}(t),\hspace*{3cm}(1)$$
where,$p$ is the car density; $f$ is the quantity responsible for the random
delay due to the say, different driving habits and road conditions. 
So one has discrete dynamical nonlinear system, which exhibit rich
dynamical behavior(see below after the presentation of two-dimensional
traffic model)\\  
Not so long ago Biham et al.(ref.2) (below simply BML) introduced a simple 
two dimensional (2D) CA model with traffic lights and studied the average 
velocity of cars as a function of their density.In that model, cars moving 
from west to east attempt to move in odd time steps and cars from south to 
north -in even time steps.There are three possible states on the square 
lattice:(i) occupied by an eastbound car; (ii)occupied by a northbound car;
(iii)vacant. In ref.8 it was underlined that the divison of time into odd and 
even time steps simulates the effect of of traffic lights.The main result of 
BML:the average velocity in the long time limit vanishes when the density 
of cars $p$ is higher than a critical value $p_{c}$.Below $p_{c}$, the 
traffic is in a moving phase, while above $p_{c}$ it is in a jamming 
phase.Numerous improvements of the basic BML model taking into account 
the effects of factors such as overpasses, faulty traffic lights, 
asymmetric distribution of cars in a homogenous lattice, and traffic 
accidents,(see ref.8 and refenences therein.)has been carried out.
In the very recent paper (ref.8) an improved mean field theory for 2D 
traffic flow with a fraction $c$ of overpass sites and with possible 
asymmetry in the distributions of cars in the two different directions is 
studied. The model in fact is the improved version of Nagatani model, ref.11.
The overpass sites can be occupied simultaneously by an eastbound car and
a northbound car, thus modelling the two-level overpasses in modern 
road systems in cities.Nagatani model deals with the isotropic distribution 
of cars with different overpass sites.It has been shown that the addition of 
overpasses enhances both the average speed of the traffic and critical 
density of cars.However the Nagatani model has some shortcomings in the sense 
that blockage  of cars due to cars moving in the same direction is not taken 
into account properly, which led to too low estimation of the concentration 
of overpasses for the transition  from jamming to moving phase at $p=1$ and 
too high estimate of the critical car density at $c=0$.
Although in this paper we will deal with the isotropic distributions 
of cars for the sake of completeness we first write the system of equations 
with asymmetry.
Let $p_{x}$ and $p_{y}$ to be the density of cars in the $x$ (eastbound) and
$y$(northbound) directions respectively.Also let $v_{x}$ and $v_{y}$ to be 
the average speeds of cars in the same directions;Let also $c$ to be 
a fraction of overpass sites.Then according to ref.8 these quantities are 
related through the following nonlinear dynamical equations:
$$v_{x}=1-(1-c)(p_{y}v_{y}^{-1}+p_{x}(v_{x}^{-1}-1))$$
$$v_{y}=1-(1-c)(p_{x}v_{x}^{-1}+p_{y}(v_{y}^{-1}-1)),\hspace*{3cm}(2)$$
From the matematical point of view the system (2)also is the nonlinear 
dynamical system.\\
It is well-known that some dynamical systems depending on the value
of systems' parameters exhibit  unpredictable,chaotic behaviour,refs.10-15.
The seminal papers (refs.12-13) induced avalanche of research works in
the theory of control of chaos in synergetics.Chaos synchronization in
dynamical systems is one of such ways of controlling chaos.
In the spirit of refs.12-13 by synchronization of two systems we mean that
the trajectories of one of the systems will converge to the same values
as the other and they will remain in step with each other.
For the chaotic systems synchronization is performed by the linking
of chaotic systems with a common signal or signals (the so-called
drivers)\\
According to refs.12-13 in the above mentioned way of chaos control
one or some of these state variables can be used as an input to drive
a subsystem that is a replica of part of the original system.In refs.12-13
it has been shown that if all the Lyapunov exponents  (or the largest
Lyapunov exponent) or the real parts of these exponents for the subsystem are 
negative then the subsystem synchronizes to the chaotic evolution of original 
system.If the largest subsystem Lyapunov exponent is not negative then as it
has been proved in ref.18 synchronism is also possible. In this case a
nonreplica system constructed according to ref.18 is used instead of replica 
subsystem.\\
The interest to the chaos synchronization in part is due to the
application of this phenomenen in secure communications, in modeling
of brain activity  and recognition processes, etc, see,references in ref. 17).
Also we should mention that this method of chaos control may result in
the improved performance (according to some criterion) of chaotic systems 
(see, e.g.ref.17).\\
In this paper for the first time (to our knowledge) we report on the possible
chaos control in the traffic flow models.\\
We will act within the algorithm proposed in ref.18.\\
Our paper is dedicated to the study of the stablization of unstable behaviors
in one dimensional and application of both replica and nonreplica
approaches  to chaos control to the 2D traffic model.\\
First we will investigate the one dimensional model.The stationary values of 
average velocity can be easily calculated from 
the equation (1).These values are:
$$v^{st}_{1}=0,v^{st}_{2}=\frac{1-(1-4(1-f)p(1-p))^{\frac{1}{2}}}{2p},\hspace*{1cm}(3)$$
The stability analysis of this stationary states show that :the $v^{st}_{1}$
is always unstable,except for $p=1$. Indeed,for this state
$$\vert\frac{v(t+1)}{v(t)}\vert =\vert (2-f)\vert >1,\hspace*{3cm}(4)$$
As $f$ changes between zero and unity.The instability condition for the 
second stationary state is:
$$\vert (1-p)^{-1}\frac{3(1-4(1-f)p(1-p))-1-2(1-4(1-f)p(1-p))^{\frac{1}{2}}}{4p} +2-f\vert >1,\hspace*{0.1cm}(5)$$
As the stability analysis show in general we have stable and unstable states
depending on the value of $p,f$. As a rule the unstable states are discarded
as unphysical ones.But nowadays due to the success of chaos control theory 
it is possible to stabilize the unstable fixed points or periodic orbits.
Below in dealing with the control of instability of fixed points in one
dimensional map we will follow the so-called proportional feedback method 
described in ref.19, which is map based variation of a method proposed 
in ref.12.Following the method  presented in ref.19 first we linearize 
the one dimensional map the one dimensional map (1) in the vicinity of 
the fixed points (or stationary states $v^{st}$):
$$v(t+1)=h(v(t)-v^{st})+v^{st},\hspace*{5cm}(6)$$
where $\vert h \vert >1$ is the slope of the map at $v^{st}$.
In the mapping (1) we have only one parameter $f$ by changing which 
one can stabilize the unstable fixed points.The positive answer to this problemvindicates the intuition that by improving drving habit,road conditions one cansolve some traffic difficulties.Of course, theoretically the regularization 
of traffic flow also could be achieved by manipulation with $p$.Having this in mind let us denote the parameters as m=(f,p).Now suppose that we change 
this parameter $m$ by small amount $\delta m$ to move the unstable fixed point $v^{st}$ without significant changing of the slope of  the map (1) $h$. In 
other words 
$$v_{t+1}(m+\delta m)=h(v_{t} - v^{st}(m+\delta m)) + v^{st}(m+\delta m),\hspace*{0.5cm}(7)$$
(in order to avoid confusion ,where necessary we write $t$ as a subscript)
where 
$$v^{st}(m+\delta m)=\delta m \frac{dv^{st}}{dm} +v^{st},\hspace*{2cm}(8)$$
Now suppose that$v_{t}=v^{st}(m)+\delta_{1} v$ ,where the second term in the 
right-hand side of this equality is much smaller than the first one.
If at this moment $m$ is changed to $m+\delta m$ such that
$v_{t+1}(m+\delta m)=v^{st}(m)$, the system state is directed to the original
unstable fixed point upon the next iteration.If $m$ is then switched back to
its original value, the system would remain at $v^{st}$ indefinitely.
The necessary variations of $m$ can  easily be determined by the formula
$$\delta m=\frac{h}{(h-1)\frac{dv^{st}}{dm}}\delta_{1} v=\frac{\delta_{1}v}{g},\hspace*{0.1cm}(9)$$
One can see easily the necessary changing of parameters to stabilise the 
unstable fixed points is proportional to the deviation of $v$ from the fixed 
point (or stationary state).That is why the method is called the proportional
-feedback one. So using the proportional feedback method can allow one to
stabilize  unstable stationary states\\
Now we will study the two dimensional case.Below, as it was underlined above 
in this paper we restrict ourselves to the isotropic case.\\
The system of equation (2)can be regarded as a mapping describing the 
time evolution of the velocity in the moving phase with $v_{x}(t+1)$ and
$v_{y}(t+1)$ on the left-hand side and $v_{x}(t)$, $v_{y}(t)$ on the 
right-hand side.\\
In the case of isotropic distribution this mapping can be written as:
$$v_{x}(t+1)=1-(1-c)(\frac{p}{2}v_{y}^{-1}+\frac{p}{2}(v_{x}^{-1}-1))=F_{1},$$
$$v_{y}(t+1)=1-(1-c)(\frac{p}{2}v_{x}^{-1}+\frac{p}{2}(v_{y}^{-1}-1))=F_{2},\hspace*{3cm}(10)$$
where $p_{x}=p_{y}=\frac{p}{2}$.\\
The system of nonlinear mapping has two steady state solutions
$$v_{\pm}=\frac{1}{2}(1+\frac{(1-c)p}{2}\pm((1+\frac{(1-c)p}{2})^{2}-4(1-c)p)^{\frac{1}{2}}),\hspace*{1cm}(11)$$
where $v=v_{x}=v_{y}$.\\
First of all we should find the condition of possible chaoticity in the 
system (10).\\
The stability of mapping is determined by the eigenvalues of the Jacobian
matrix of the nonlinear mapping (10).
$$J=\frac {\partial(F_{1},F_{2})}{\partial(v_{x}(t),v_{y}(t))}, \hspace*{2cm}(12)$$
It can be seen easily the eigenvalues of the Jacobian matrix is calculated 
by the following equation:
$$\lambda^{2}-\lambda (1-c)\frac{p}{v^{2}}=0,\hspace*{6cm}(13)$$
From here we obtain easily that 
$$\lambda_{1}=0,$$
$$\lambda_{2}=(1-c)\frac{p}{v^{2}},\hspace*{5cm}(14)$$
In the last expression while calculating $\lambda$ we use
the steady state solutions (10). This simplification is justified
at least for systems whose chaotic behavior has arisen out of stability
of fixed points, see ref.15, also ref.20-21.\\ 
The mapping will exhibit chaotic behaviour, if the absolute values of $\lambda$
exceed unity.
$$\vert\lambda_{2}\vert>1,\hspace*{7cm}(15)$$.
 As the initial or original nonlinear mapping is symmetric over $v_{x}$ 
and $v_{y}$ considering only one of these variables as a driver 
will be sufficient.Take for the definiteness $v_{x}$ variable as a driver.
Then the replica subsystem(with the superscript"r") can be written as follows:
$$v^{r}_{y}(t+1)=1-(1-c)(\frac{p}{2v_{x}(t)}+\frac{p}{2}(\frac{1}{v^{r}_{y}(t)}-1))=H,\hspace*{1cm}(16)$$
Then the Lyapunov exponent can be calculated as follows,ref.18
$$\Lambda=\ln\frac{\partial H}{\partial v^{r}_{y}}=\ln(1-c)\frac{p}{2}\frac{1}{v^{2}},\hspace*{3cm}(17)$$.
For the chaos control (to be more specific for the synchronization  of the 
evolution of the response system to the chaotic evolution of the initial 
nonlinear mapping when time goes to infinity) it is required that
$$ \Lambda <0,\hspace*{9cm}(18)$$.
It will take place, if the following condition is satisfied:
$$(1-c)\frac{p}{2v^{2}}<1,\hspace*{7cm}(19)$$.
Thus we have two inequalities:(19) and (15).If these inequalities 
do not contradict each other, then the replica approach allows us to 
perform chaos control.
We see that chaos control within replica approach is realizable if
$$1<(1-c)p\frac{1}{v^{2}}<2,\hspace*{5cm}(20)$$
Although the restriction (20) is very severe in the sense that 
diapason of changing of traffic flow models parameters such as $c$, $p$
could be very narrow.Nevertheless, as the analysis of the data 
presented in ref.8 shows that chaos synchronization would be possible within 
the replica approach.If this approach fails, we can apply nonreplica 
one to achieve our goal,as it was underlined above.\\
Now suppose that our attempts to perform chaos control failed within 
replica approach.\\
In this case we can try nonreplica approach.
According to ref.18, within nonreplica approach we can use the following 
nonreplica response system(with the superscript " nr"):
$$v^{nr}_{x}(t+1)=1-(1-c)(\frac{p}{2}(v^{nr}_{y})^{-1}+\frac{p}{2}(v_{x}^{-1}-1))+\alpha(v^{nr}_{x}-v_{x})=F_{3},$$
$$v^{nr}_{y}(t+1)=1-(1-c)(\frac{p}{2}v_{x}^{-1}+\frac{p}{2}((v^{nr}_{y})^{-1}-1))+\beta(v^{nr}_{x}-v_{x})=F_{4},\hspace*{3cm}(21)$$
where $\alpha$ and $\beta$ are the arbitrary constants.\\ 
Here again as in the previous case we consider $v_{x}$ dynamical variable
as the driver.\\ 
The Lyapunov exponents are the eigenvalues of the Jacobian:
$$J=\frac{\partial(F_{3},F_{4})}{\partial(v^{nr}_{x}(t),v^{nr}_{y}(t))},\hspace*{2cm}(22)$$
From (21) we easily establish that these exponents are solutions to the 
following equation:
$$\lambda^{2}-\lambda (\alpha + (1-c)\frac{p}{2}v^{-2})-\beta (1-c)\frac{p}{2}v^{-2}=0,\hspace*{2cm}(23)$$
Here $v$ is the steady state solution of the original mapping.
As it can be seen from (23) the roots of this equation $\lambda_{1}$
and $\lambda_{2}$ satisfy the relationships:
$$\lambda_{1} + \lambda_{2}=\alpha + (1-c)\frac{p}{2}v^{-2},$$
$$\lambda_{1}\lambda_{2}=-\beta (1-c)\frac{p}{2}v^{-2},\hspace*{7cm}(24)$$
Remind that our aim is to satisfy the conditions  
$\vert \lambda_{1}\vert <1$ and $\vert \lambda_{2}\vert <1$. 
Due to the arbitraryness of the constants $\alpha$ and $\beta$ this can be done
easily.\\
Up to now while performing chaos synchronization we have taken the advantage ofusing the presence of driving variables explicitly.Our calculations show that 
chaos synchronization is also reachable in case of parameter perturbation method,ref.22. Namely, we have shown that by changing the fraction of overpasses onecan make the absolute values of $\lambda_{1,2}$ less than unity.
Indeed, by assuming the following change for $c$ : 
$$c=c_{1}-\alpha_{c}(v_{y}- v_{y_{ap}}),\hspace*{5cm}(25)$$
(where:$c_{1}$ is the nominal value for the $c$ in the original two dimensional
model;$\alpha_{c}$ is the control coefficient to be found;$v_{y}$ and
$v_{y_{ap}}$ are are response system and drive system orbits,respectively.)
after lengthy calculations we obtain that for the isotropic case  
much sought eigenvalues are:
$$\lambda_{1}=0,$$
$$\lambda_{2}=2(1-c_{1})\frac{p}{2} v^{-2}_{ap}+\frac{p}{2}\alpha_{c} (1-2v^{-1}_{ap}),\hspace*{0.5cm}(26)$$
Thus we have the real possibility to satisfy the condition for the chaos control:$\vert \lambda_{2} \vert <1$.\\ 
Now let us make some estimations based on our approach.According to ref.8
for the fraction of overpasses $c=0.5$ the critical value of car density 
when jam occurs 
$$p_{cr}=\frac{6-32^{\frac{1}{2}}}{1-c}=0.686,\hspace*{1cm}(27)$$
Then for $p>p_{cr}$ jam phase takes place with v close to zero.It means 
that for these values of $c,p$  instability conditions hold.
Now we will use the formulae (25)and (26)to resolve the jamming problem 
by changing $c$.Of course, while conducting calculations we should keep in 
mind that maximal values for $c, v$ are unity and the jamming phase could 
be avoided by increasing $c$.From the condition $\vert \lambda_{2}\vert >1$
(formula (26)) for the extreme case $v_{yap}=1$ we obtain that 
$-1.9<\alpha_{c} <3.9$.Take for definiteness
$\alpha_{c}=-1.5$.Further, having in mind that when synchronization takes place
$v_{y}$ is very close to $v_{yap}$, therefore assuming $v_{y}-v_{yap}=0.1$
from the equation (25) we derive  new value for the fraction of overpasses 
capable of resolving the traffic jamming:$c=0.65$ which is very close to the 
value of $c$,$c=0.657$ when there is no jamming at all;that is $p_{cr}=1$.,ref.8.
Of course much depends on the degree of accuracy for synchronization between
$v_{y}$ and $v_{yap}$.But,in any case we can say safely that our results do notcontadict to the conventional visdom and gives the right direction of action\\ In conclusion in this work we pointed out to the possibility of the
stabilization of the unstable stationary states in one of one dimensional
model with random delay. Also we have investigated the possibilty of chaos 
control in one of two dimensional mapping in traffic flow within replica, 
nonreplica and parameter change approaches.As we indicated above nonreplica approach has 
advantages over the replica approach.One of them is the possibility to make 
Lyapunov exponents not only negative, but also larger in magnitude.This is 
very important from the application point of view, as the time required to 
achieve synchronization depends on the value of the largest Lyapunov exponent, ref.1-16, 18.\\
Speaking about the study of 2D  CA traffic flow models one has to mention
that although 2D CA models less representative (in comparison with 1D  
rule-184 CA) of real traffic flow, however they may be applicable 
to abstract traffic problems such as data packets in computer networks,ref.23.
Besides, 2D  CA models may be useful from the viewpoint of complex behavior 
in deterministic dynamics, ref.24-26.\\ 
Acknowledgments\\
E.M.Shahverdiev thanks JSPS for the Fellowship.\\

\end{document}